\documentclass[prd,twocolumn,showpacs,floats,floatfix,nofootinbib,superscriptaddress]{revtex4}
\usepackage{dcolumn} 
\usepackage{amssymb}
\usepackage{amsmath}
\usepackage{natbib}
\usepackage{bm} 
\usepackage{graphicx}
\usepackage{amsfonts}
\usepackage{txfonts} 
\usepackage{bbm}
\usepackage{bbold}
\newcommand{\be}{\begin{equation}}
\newcommand{\ee}{\end{equation}}
\newcommand{\bea}{\begin{eqnarray}}
\newcommand{\eea}{\end{eqnarray}}
\newcommand{\beq}{\begin{equation}}
\newcommand{\eeq}{\end{equation}}
\newcommand{\beqar}{\begin{eqnarray}}
\newcommand{\eeqar}{\end{eqnarray}}
\newcommand{\beqars}{\begin{eqnarray*}}
\newcommand{\eeqars}{\end{eqnarray*}}
\newcommand{\bc}{\begin{center}}
\newcommand{\ec}{\end{center}}
\newcommand{\ben}{\begin{enumerate}}
\newcommand{\een}{\end{enumerate}}
\newcommand{\bit}{\begin{itemize}}
\newcommand{\eit}{\end{itemize}}
\newcommand{\bw}{\begin{widetext}}
\newcommand{\ew}{\end{widetext}}
\newcommand{\si}[1]{{\scriptscriptstyle{#1}}}
\newcommand{\ex}{\mbox{e}}
\newcommand{\dd}{\mbox{d}}
\def\spose#1{\hbox to 0pt{#1\hss}}

\def\lta{\mathrel{\spose{\lower 3pt\hbox{$\mathchar"218$}}
\raise 2.0pt\hbox{$\mathchar"13C$}}}
\def\gta{\mathrel{\spose{\lower 3pt\hbox{$\mathchar"218$}}
\raise 2.0pt\hbox{$\mathchar"13E$}}}

\def\setR{\mathbb{R}}
\def\setC{\mathbb{C}}

\def\Dalembert{\kern1pt\vbox{\hrule height 1.2pt\hbox{\vrule width 1.2pt\hskip 3pt 
\vbox{\vskip 6pt}\hskip 3pt\vrule width 0.6pt}\hrule height 0.6pt}\kern1pt} 
\newcommand{\ie}{\textsl{i.e., }}

\newcommand{\mrm}[1]{\mathrm{#1}}
\newcommand{\mcl}[1]{\mathcal{#1}}

\newcommand{\real}{\mathbbm{R}}

\begin{document}

\title{Nonabelian Bosonic Currents in Cosmic Strings}

\author{Marc Lilley} \email{lilley@iap.fr}
\affiliation{${\cal G}\setR\varepsilon\setC{\cal O}$ --
Institut d'Astrophysique de
Paris, UMR7095 CNRS, Universit\'e Pierre \& Marie Curie, 98 bis
boulevard Arago, 75014 Paris, France}
\affiliation{Theoretical and Mathematical Physics Group, Centre for
  Particle Physics and Phenomenology, Louvain University, 2 Chemin du
  Cyclotron, 1348 Louvain-la-Neuve (Belgium)}

\author{Fabrizio Di Marco} \email{Fabrizio.DiMarco@icranet.org}
\affiliation{ICRANet, Piazza della Repubblica 10, 65122 Pescara, Italy}

\author{J\'er\^ome Martin} \email{jmartin@iap.fr}
\affiliation{${\cal G}\setR\varepsilon\setC{\cal O}$ --
Institut d'Astrophysique de
Paris, UMR7095 CNRS, Universit\'e Pierre \& Marie Curie, 98 bis
boulevard Arago, 75014 Paris, France}

\author{Patrick Peter} \email{peter@iap.fr}
\affiliation{${\cal G}\setR\varepsilon\setC{\cal O}$ --
Institut d'Astrophysique de
Paris, UMR7095 CNRS, Universit\'e Pierre \& Marie Curie, 98 bis
boulevard Arago, 75014 Paris, France}

\begin{abstract}
A nonabelian generalization of the neutral Witten current-carrying
string model is discussed in which the bosonic current-carrier
belongs to a two dimensional representation of SU(2). We find that the
current-carrying solutions can be of three different kinds: either
the current spans a U(1) subgroup, and in which case one is left with
an abelian current-carrying string, or the three currents are all
lightlike, travelling in the same direction (only left or right
movers). The third, genuinely nonabelian situation, cannot be handled
within a cylindrically symmetric framework, but can be shown to
depend on all possible string Lorentz invariant quantities that can
be constructed out of the phase gradients. 
\end{abstract}
\date{\today}
\maketitle

\section{Introduction}

Topological cosmic strings or superstrings of cosmological size are
one-dimensional extended objects which are believed to have been
formed in the early phases of cosmological evolution.  They are of
considerable interest because they may offer a observable window on
the high energy physics of the primordial universe, \ie at grand
unified scales.

\par

Topological strings are produced in phase transitions associated with
spontaneous symmetry breaking.  This is the standard Kibble
mechanism~\cite{kibble1,kibble2}. Almost all supersymmetric grand
unified theories in which hybrid
inflation~\cite{hybrid2,Copeland:1994vg,Dvali:1994ms} can be realized
lead to the formation of topological
strings~\cite{rachel1,rachel2,JRS03,john}. Besides, most classes of
superstring compactification lead to a spontaneous breaking of a
pseudo-anomalous U$(1)$ gauge symmetry producing local cosmic strings
\cite{BDP98}. Such strings also form in the case where the Higgs field
has a non-minimal kinetic term~\cite{Babichev:2008qv}.

\par

The simplest kind of topological string is the Nambu-Goto string which
is described by the Nambu-Goto action~\cite{GN1,GN2}.  The Nambu-Goto
action is the worldsheet formulation counterpart of a field theory
description in which the string arises as a solitonic solution of the
abelian Higgs model~\cite{NO}.  Such a string has no internal
structure and is described entirely in terms of a worldsheet
Lagrangian and the tension per unit length of the string.

\par

Most observational signatures in the gravitational sector expected
from topological strings have been derived and simulated numerically
for Nambu-Goto strings.  There are five main possible observational
effects (see~\cite{VilenkinBook,PPJPU} and references therein): beamed
gravitational wave bursts from kinks and particle acceleration;
deflection, gravitational lensing effects and multiple image effects;
Doppler shifting effects; background gravitational radiation from
string loops; and string effects in the cosmic microwave background.
The existence of kinks along the strings has been shown to occur also
for current-carrying strings~\cite{3D} and the electromagnetic effects
of such strings, which are absent in the simpler Nambu-Goto string,
have been investigated.  An especially interesting observational
consequence of the presence of cosmic string networks in the early
universe potentially because it is susceptible to be detected in the
cosmic microwave background is the Gott-Kaiser-Stebbins
effect~\cite{Gott:1984ef,Kaiser:1984iv}.  This effect consists in a
temperature shift that is due to the gravitational lensing of photons
passing near a moving source.

\par

Cosmic superstrings are formed by tachyon condensation at the end of
brane inflation~\cite{Sarangi:2002yt,Jones:2003da}.  The tachyons are
complex scalars [with a local U(1) gauge symmetry] identifiable with
the ground state open string modes of the Neveu-Schwarz sector that
end on coincident non-BPS branes and
antibranes~\cite{Green:1994iw,Banks:1995ch,Green:1996um,Lifschytz:1996iq}.
There exist associated gauge fields living on the brane and antibrane
so that there exists a U(1)$\times$U(1) symmetry on the
brane-antibrane configuration.  A first linear combination of the
U(1)'s is higgsed~\cite{Dvali:2002fi,Dvali:2003zj} leading to the
appearance of a first kind of cosmic superstrings that are D
$p$-branes with $p-1$ dimensions compactified~\cite{Majumdar:2002hy}.
In type IIB superstring theory, and given a spacetime manifold $\mcl
M$, such stable $p$-branes, can, for example, be obtained by
considering a $p+2$ brane-antibrane pair stretching over a submanifold
$\real^{p+3}\subset \mcl M$.  The $p+2$ brane-antibrane pair will
annihilate unless a topological obstruction exists.  This obstruction
can be obtained from
K-theory~\cite{Sen:1998tt,Witten:1998cd,Olsen:1999xx}.  A second
linear combination of the U(1)'s leads to the formation of
F-strings~\cite{Dvali:2002fi,Dvali:2003zj}.

\par

All these types of strings have until recently been considered as
structureless, so their dynamics is given by the Nambu-Goto
action. Numerical simulations of networks (see \cite{FRSB08} and
references therein) of such strings have been produced with the result
of scaling, a property thanks to which the string network never comes
to dominate the Universe evolution, but neither are the string
completely washed out of the Universe, so their effect, however small,
is still detectable.

\par

The Nambu-Goto string can be generalized to the case of a string with
internal structure.  Such a string can be obtained by including a
coupling of the string forming Higgs field to additional (bosonic or
fermionic, with global or local, abelian, or nonabelian symmetry)
fields in the theory.  In part of the parameter space, these fields
condense onto the string (the symmetry gets broken) leading to the
appearance of currents on the worldsheet in the form of Goldstone
bosons propagating along string~\cite{Witten:1984eb}.  In such a case,
the current-carrying string can be described using a worldsheet
Lagrangian and a nontrivial equation of state relating the tension per
unit length to the energy density of the
string~\cite{formal3,carterPLB89,formal2,formal4}; the actual form of
this equation of state was discussed numerically
\cite{neutral,enon0,NoSpring} and analytically \cite{models}. The
presence of currents on the worldsheet modifies only slightly the
gravitational properties of the long strings
\cite{GravStringPuyPP,GP94}, but it also halts cosmic string loop
decay caused by dissipative effects, thereby yielding new equilibrium
configurations \cite{2D,3D} named
vortons~\cite{vortons0,vortons1,vortons2,vortons3,vortons4,vortons5}. Those
can potentially change drastically the cosmological network evolution,
at the point of ruling such strings out.

\par

Although the current-carrying property of cosmic strings is in fact
fairly generic \cite{lowmass,DP95,PGdA03}, a possibility that has,
until now, been completely disregarded is that for which the string
would be endowed not only with many currents \cite{LPXcoupled}, but
also with currents of a nonabelian kind, as is to be expected in most
grand unified theories. This natural extension of the Witten idea
leads to numerous new difficulties, as in particular the internal
degrees of freedom manifold is intrinsically curved, so that a local,
flat, description of the string worldsheet manifold, turns out to be
inappropriate \cite{BCIII,BCIV}. This paper is devoted to the specific
task of obtaining the equivalent microscopic structure of a nonabelian
current-carrying cosmic string.

\par

To do so, we restrict attention to the global situation in which, in a
way similar to the so-called neutral Witten model \cite{neutral}, we
wish to capture the essential internal dynamics of the string without
the undue complication of adding extra gauge vector fields. In the
case of an abelian current, it was indeed shown that these
contributions, although of potential great cosmological relevance
(see, e.g. Ref.~\cite{boucleem} and references therein), can however
be treated in a perturbative way, not modifying in any essential way
the actual microscopic structure \cite{enon0}.  We therefore assume,
as a toy model, a U$(1)$ Higgs model whose breaking leads to the
existence of the strings themselves, coupled to an SU$(2)$ doublet
through a scalar potential with parameters ensuring a condensate. We
first describe the fields and notation, derive their dynamical
equations in full generality, and then discuss the condensate
configuration. After having recovered the abelian cases as particular
solutions of the general nonabelian situation, we concentrate on the
strictly nonabelian solutions. We obtain an exact configuration,
called trichiral, and show how this model makes explicit the
obstruction theorem first obtained by Carter \cite{BCIII,BCIV}. We
then derive the stress energy tensor and its eigenvalues, namely the
energy per unit length and tension, and show that they depend on all
the possible two-dimensional Lorentz invariants that can be
constructed from the phase gradients (and second derivatives) of the
angular variables in the internal space. We conclude by discussing the
possible cosmological consequences of this new category of objects.

\section{Fields content}
\label{fieldcontent}

The simplest nonabelian current-carrying string model that can be
written down is that in which a U$(1)$ symmetry is spontaneously
broken by means of a scalar complex Higgs field $\phi$, itself coupled
to $\bm{\Sigma}$, a scalar field belonging to an arbitrary
representation of a nonabelian group $G$.  The string-forming action
stems from the Higgs Lagrangian
\begin{equation}
\mcl{L}_{_\mathrm{S}} = -D_{\mu}{\phi}^\star D^{\mu}{\phi}
- \frac{1}{4}C_{\mu \nu} C^{\mu \nu} - V _{_\mathrm{H}} \left( \phi \right),
\label{Higgs}
\end{equation}
where
\begin{equation}
C_{\mu \nu} =\nabla_{\mu} C_{\nu}- \nabla_{\nu} C_{\mu}
\end{equation}
and the U$(1)$ covariant derivative is expressed in terms of the
U$(1)$ gauge field $C_\mu$ as
\begin{equation}
D_{\mu}\phi = \nabla_{\mu} \phi + i q C_{\mu} \phi,
\end{equation}
where $q$ is the charge. $V _{_\mathrm{H}} $ can be chosen without
lack of generality as the Higgs symmetry breaking potential, namely
\begin{equation}
V _{_\mathrm{H}} = \displaystyle\frac{\lambda_\phi}{4}
\left( \left\vert\phi \right \vert^2 -\eta^2\right)^2,
\label{HiggsPotential}
\end{equation}
with $\lambda_\phi$ a coupling constant and $\eta$ the Higgs vacuum
expectation constant (vev) at infinity.

\par

The current part of the Lagrangian reads
\begin{equation}
\mcl{L}_{_\mathrm{C}} =
-\left( \partial_{\mu}{\bm{\Sigma}}\right)^\dagger
\cdot \partial^{\mu}\bm{\Sigma} 
- V_{_\mathrm{C}}\left(\bm{\Sigma}\right),
\label{LC}
\end{equation}
where $\bm{\Sigma}$ transforms according to a yet arbitrary
representation of the global invariance group whose structure
constants we write as $f^a_{\ bc}$; these are defined through the
commutation relations for $\{T^a\}$, the algebra of $G$, namely
\begin{equation}
\left[ T^a, T^b \right] = i f^c_{\ ab} T_c.
\label{algebra}
\end{equation}
In Eq.~(\ref{algebra}) and in the following, the group indices are
denoted by latin smallcap letters $a,b,\cdots=1,\cdots,N$ which run to
$N$, the group dimension. The potential appearing in the current
action is the self-interacting potential chosen as
\begin{equation}
V_{_\mathrm{C}}\left(\bm{\Sigma}\right) = \pm m_\sigma^2 
\bm{\Sigma}^\dagger\cdot\bm{\Sigma} + \lambda_\sigma 
\left(\bm{\Sigma}^\dagger\cdot\bm{\Sigma}\right)^2,
\label{PotSig}
\end{equation}
thus introducing the vacuum mass and self-interaction constants
$m_\sigma$ and $\lambda_\sigma$. In Eq.~(\ref{PotSig}), we have
introduced a sign parameter which accounts for the possibility that
SU$(2)$ is broken ($-$) or unbroken ($+$) far from the string
core. The first possibility is usually not taken into account when one
considers the Witten model since in that case, one has in mind that
the condensate depicts electromagnetism, which is obviously unbroken
far from the string. In the nonabelian case however, it is reasonable
to assume a broken symmetry far from the string as well, in particular
if one is to identify this symmetry with that of the electroweak
phenomenology.

\par

The total action of the system can be written as
\begin{equation}
\label{actiontot}
{\cal L}={\cal L}_{_\mathrm{S}}+{\cal L}_{_\mathrm{C}}-V_\mathrm{int}\, ,
\end{equation}
where the interaction term couples the two scalar fields
$V_\mathrm{int}\left(\phi,\bm{\Sigma}\right)$. This potential, again
for illustrative purposes below, shall be taken as the most general
renormalizable one, namely
\begin{equation}
V_\mathrm{int}\left(\phi,\bm{\Sigma}\right) = f \left( \left \vert 
\phi \right \vert ^2 - \eta^2\right) 
\bm{\Sigma}^\dagger \cdot \bm{\Sigma},
\label{Vint}
\end{equation}
with a positive coupling constant $f$ to ensure vacuum stability. The
vacuum far from the string therefore depends on the representation
$\bm{\Sigma}$ belongs to. The microscopic parameters that allow for a
condensate to form are similar to those of the abelian current case;
they have been discussed in particular in Ref.~\cite{neutral}.

\section{Field equations}
\label{fieldeqs}

Having specified the field content and the action of the system, one
can now derive the corresponding equations of motion. The equations of
motion of the system consisting of the string-forming fields $\phi$
and $C_\mu$ and the current carrier $\bm{\Sigma}$ are
\begin{equation}
\nabla_{\mu}\nabla^{\mu} \phi + 2iq C^{\mu}\nabla_{\mu}\phi + i q
\phi \nabla_{\mu} C^{\mu} - q^2 C^{\mu} C_{\mu} \phi - \frac{\partial
V}{\partial \phi^*} = 0,
\label{eqphi}
\end{equation}
for the string-forming Higgs field,
\begin{equation}
\nabla_{\mu} C^{\mu \nu} - iq (\phi \nabla^{\nu}
\phi^* - \phi^* \nabla^{\nu}\phi) - 2 q^2 C^{\nu} |\phi|^2 = 0
\label{eqF}
\end{equation} 
for the associated U$(1)$ gauge field, and
\begin{equation}
\Dalembert \bm{\Sigma} = \frac{\partial V}{\partial \bm{\Sigma}^\dagger}
\label{eqsigma}
\end{equation}
for the current carrier.

\par

The energy-momentum tensor of the system is given by the usual
relation
\begin{equation}
T_{\mu \nu}\equiv g_{\mu \nu}{\mathcal{L}} 
- 2 \frac{\delta \mathcal{L}}{\delta g^{\mu\nu}},
\end{equation}
and can be decomposed into a scalar and a vector part, namely
\begin{equation}
T_{\mu \nu} = T^\mathrm{s}_{\mu \nu} + T^\mathrm{v}_{\mu \nu},
\end{equation}
\noindent where
\begin{eqnarray}
T^\mrm{s}_{\mu \nu}&=& D_{(\mu} \phi^* D_{\nu)} \phi 
- g_{\mu \nu } D_\gamma \phi^* D^\gamma \phi
\nonumber\\
& & + \partial_{(\mu} \bm{\Sigma}^\dagger \cdot \partial_{\nu)} 
\bm{\Sigma} - g_{\mu \nu } 
\left( \partial_\gamma \bm{\Sigma}\right)^\dagger \cdot 
\partial^\gamma \bm{\Sigma}
\nonumber\\
& & - g_{\mu \nu }  V\left(\phi, \bm{\Sigma}\right),
\label{energy_scalar}
\end{eqnarray}
with parentheses denoting symmetrization of the indices, i.e.,
$S_{(\alpha\beta)}\equiv S_{\alpha\beta} + S_{\beta\alpha}$, and
\begin{equation}
T^{\mrm{v}}_{\mu\nu} = - \left( F_{\mu\alpha} F^\alpha_{\ \nu} +
\frac{1}{4} g_{\mu \nu } F_{\alpha \beta}F^{\alpha\beta} \right)\, ,
\label{energy_vector}
\end{equation}
where we have defined $F_{\alpha \beta }\equiv \partial
_{\alpha}C_{\beta }-\partial _{\beta}C_{\alpha }$.

\par

From this stress-energy tensor and the field equations, we shall now
derive the full microscopic structure of the system.

\section{The condensate}
\label{condensate}

Having derived the most general form of the equations of motion, we
now turn to the specific situation where an straight, infinitely long,
cosmic string is present. A typical vortex solution aligned along the
$z$ axis in polar coordinates $r$ and $\theta$ is then given by the
Nielsen-Olesen ansatz
\begin{equation}
\phi = \varphi(r) \ex^{i n \theta} \ \ \ \hbox{and} \ \ 
\ C_\mu = C_\theta (r) \delta_\mu^\theta,
\label{NO}
\end{equation}
where $n\in\mathbb{Z}$. Although the specific form of the potential is
irrelevant for most of what follows, the shape (\ref{HiggsPotential}),
being the most general renormalizable function satisfying this
constraint, is used in the numerical illustrations below. Inserting
the above ansatz into the equations of motion, Eq.~(\ref{eqphi}) takes
the form
\begin{equation}
\frac{\dd^2\varphi}{\dd r^2}+\frac{1}{r}\frac{\dd \varphi}{\dd r} =
\frac{Q^2}{r^2}\varphi+ 
\frac{\partial V}{\partial\varphi},
\label{varphi}
\end{equation}
while Eq.~(\ref{eqF}) becomes
\begin{equation}
\frac{\dd^2Q}{\dd r^2}-\frac{1}{r}\frac{\dd Q}{\dd r}=2q^2 Q\varphi^2,
\label{Q}
\end{equation}
where we have defined $Q\equiv n+qC_{\theta}$. In Eq.~(\ref{varphi}),
the last term of the r.h.s involves not only the derivative of the
self-interaction potential $V_{_\mathrm{H}}$, but also that of the
coupling term $V_{\mathrm{int}}$, so that this equation also depends
on the SU$(2)$ doublet amplitude. It is through this ``backreaction''
term that the string itself is affected by the presence of the
current.

\par

Let us now discuss in more detail the form of the current-carrier
scalar field $\bm{\Sigma}$. Our goal is to find the most general
ansatz for $\bm{\Sigma}$ in cylindrical coordinates. The case where
$G=\mathrm{U}(1)$ represents the usual so-called superconducting
string model originally introduced by Witten~\cite{witten1}. In this
particular case, $\Sigma$ is a complex field vanishing in vacuum,
i.e. far from the string. Its coupling with the string-forming Higgs
field yields an instability in the vortex core leading to a
condensate: far from the string, in vacuum, where the Higgs field is
equal to its vev $|\phi|=\eta$, the interaction term $V_\mathrm{int}$
vanishes so that $\Sigma$ must vanish. The string location, defined as
the set of points where $\phi=0$, however, is no longer vacuum-like
from the point of view of $\Sigma$, and indeed the parameters of the
potential (\ref{Vint}) can be chosen~\cite{neutral,enon0} such that
$\Sigma$ does not vanish inside the vortex.

\par

One can pick a specific gauge in which $\Sigma$ is real, $\Sigma =
\sigma(r)$ say, depending only on the distance to the string, with
$\sigma\in\setR$ and $\lim_{r\to\infty}\sigma(r)=0$, and generate all
the solutions by applying a gauge transformation, in this case a
phase. The full solution then reads
\begin{equation}
\Sigma = \ex^{i\psi(z,t)T}\sigma(r),
\label{sigU1}
\end{equation}
where the phase transformation can now depend on the worldsheet
internal coordinates and we did not take into account a possible
dependence in the external coordinates. In Eq.~(\ref{sigU1}), we have
written explicitly the generator of the U(1) translation as $T$, even
though it is not necessary in this simplifying case for which the
scalar field is a mere singlet under this extra U(1); note that this
could be different if $\Sigma$ were belonging to the representation of
a larger group containing this U(1).

\par

Written in the form (\ref{sigU1}) with the generator, the solution is
easily generalizable to the nonabelian case. We again choose a gauge in
which $\bm{\Sigma} = \bm{\sigma} (x^\perp)$, with $\bm{\sigma}$ in the
desired representation but depending only on the external coordinates
$x^\perp $ (in practice the radial distance $r$), and produce the full
solution by exponentiation of the generators $T_a$ as
\begin{equation}
\bm{\Sigma} = \ex^{i\psi^a\left(\xi\right)T_a}\bm{\sigma}(x^\perp),
\label{sigG}
\end{equation}
where the functions $\psi_a$ a priori depend on the internal
coordinates $\xi$ only. As it turns out however~\cite{BCIII,BCIV}, in the
more general case of a nonabelian symmetry, the fields $\psi_a$ live
on a curved manifold which cannot, in general, be smoothly projected
on the flat manifold describing the string worldsheet. As a result,
one must assume that the fields $\psi_a$ depend on all embedding
coordinates.

\par

The form (\ref{sigG}) is not, unfortunately, directly usable, as the
derivative of the group element is not easy to handle. Indeed, for a
noncommuting algebra, one has
\begin{equation}
\partial_\mu U = i \partial_\mu\bm{\psi} \cdot\int_0^1 U(1-p) \bm{T} U(p) \dd p
\not= i\partial_\mu \bm{\psi} \cdot \bm{T} U,
\label{dU}
\end{equation}
where $U(p) \equiv \exp (i p\bm{\psi}\cdot\bm{T})$ and $U\equiv \exp
(i\bm{\psi}\cdot\bm{T}) = U(1)$ and, the last relation becoming an
equality in the abelian case. Restricting attention to SU(2) however,
allows simple calculations to be carried out completely since one then
has the useful relation
\begin{equation}
\ex^{i\alpha \bm{n}\cdot\bm{\tau}} = \cos\alpha \mathbb{1}+ i
\bm{n}\cdot\bm{\tau} \sin\alpha, \ \ \ \hbox{with} \ \ \ n_a n^a =1,
\label{Usig}
\end{equation}
between the Pauli matrices $\tau^a$, generators of SU(2), and their
exponentiated form. We therefore restrict attention to a scalar field
belonging to the representation $\bm{2}$ of SU(2), i.e. a doublet, and
thus assume in what follows that the current-carrier takes the form
\begin{equation}
\bm{\Sigma} = \left( \cos\alpha \mathbb{1}+ i \bm{n}\cdot\bm{\tau} \sin\alpha
\right) \sigma g, \ \ \ \ \hbox{with} \ \ \ g\equiv
\frac{1}{\sqrt{2}}\left(\begin{array}{c}0\cr 1\end{array} \right).
\label{Sigma}
\end{equation}
Notice that Eq.~(\ref{varphi}), together with the assumption of a
potential depending only on the amplitude
$\bm{\Sigma}^\dagger\cdot\bm{\Sigma} = \frac12\sigma^2$, shows that
$\sigma = \sigma(r)$ only. But, as already mentioned above, the angle
$\alpha $ and the normalized vector $n^a$ a priori depend on all the
coordinates. 

\par

With the form (\ref{Sigma}) for the scalar field, the variation of
the potential is
\begin{equation}
\frac{\partial V}{\partial
  \bm{\Sigma}^\dagger} = \frac12 \frac{\partial V}{\partial \sigma}
\left( \cos\alpha \mathbb{1}+ i \bm{n}\cdot\bm{\tau} \sin\alpha \right) g,
\label{dVdsigma}
\end{equation}
which provides the equation of motion through
Eq.~(\ref{eqsigma}). Indeed, projecting this equation of motion on the
identity of SU(2) yields
\begin{equation}
\Delta \sigma - \left[ \left(\partial\alpha\right)^2 + \tan\alpha
\Dalembert \alpha\right] \sigma - 2 \tan
\alpha\partial\alpha\cdot \partial\sigma = \frac12 \frac{\partial
V}{\partial\sigma}, 
\label{Id}
\end{equation}
while the projection on the Pauli matrices $\tau^a$ leads to
\begin{widetext}
\begin{equation}
n^a\left\{ \Delta\sigma+
\left[\frac{\Dalembert\alpha}{\tan\alpha}-\left(\partial\alpha\right)^2
\right] \sigma
+2\frac{\partial\alpha\cdot\partial\sigma}{\tan\alpha}\right\} +
2\left( \partial\sigma +
\frac{\sigma\partial\alpha}{\tan\alpha}\right) \cdot \partial n^a +
\sigma \Dalembert n^a = \frac12 \frac{\partial V}{\partial \sigma} n^a,
\label{taua}
\end{equation}
\end{widetext}
which in turn implies, upon projection on $n_a$, recalling this vector
to be normalized to unity, that
\begin{equation}
\Delta \sigma - \left[ \left(\partial\alpha\right)^2 -
\frac{\Dalembert \alpha}{\tan\alpha} -n_a \Dalembert n^a
\right] \sigma + 2
\frac{\partial\alpha\cdot \partial\sigma}{\tan\alpha} = 
\frac12 \frac{\partial
V}{\partial\sigma}.
\label{tauana}
\end{equation}
This last equation can be used in order to simplify
Eq.~(\ref{taua}). Indeed, inserting Eq.~(\ref{tauana}) into
Eq.~(\ref{taua}), one obtains
\begin{equation}
\Dalembert n^a+2\left( \frac{\partial\sigma }{\sigma}+
\frac{\partial\alpha}{\tan\alpha}\right) \cdot \partial n^a
-\left(n_b\Dalembert n^b\right)n^a=0\, ,
\label{tauasimple}
\end{equation}
which provides a clean equation for the evolution of the vector
$n^a$. Note also that Eqs.~(\ref{Id}) and (\ref{tauana}) can be
combined to provide a dynamical equation for the angle $\alpha$,
namely
\begin{equation}
\Dalembert \alpha + \frac{2}{\sigma} \partial\sigma\cdot\partial \alpha +
\sin\alpha \cos \alpha \left( n_a \Dalembert n^a\right) =0,
\label{alphadyn}
\end{equation}
and the profile of the condensate then satisfies
\begin{equation}
 \Delta \sigma - \left[ \left(\partial\alpha\right)^2- \left(n_a \Dalembert
     n^a\right) \sin^2\alpha\right] \sigma = \frac12 \frac{\partial
   V}{\partial\sigma}, 
\label{profile}
\end{equation}
which generalizes the abelian case by inclusion of the nonlinear
term. At this stage, Eqs.~(\ref{tauasimple}), (\ref{alphadyn})
and~(\ref{profile}) are the equations that one needs to solve in order
to determine $\sigma $, $\alpha $ and $n^a$.

\par

\begin{figure*}
\hskip-5mm\includegraphics[scale=.47]{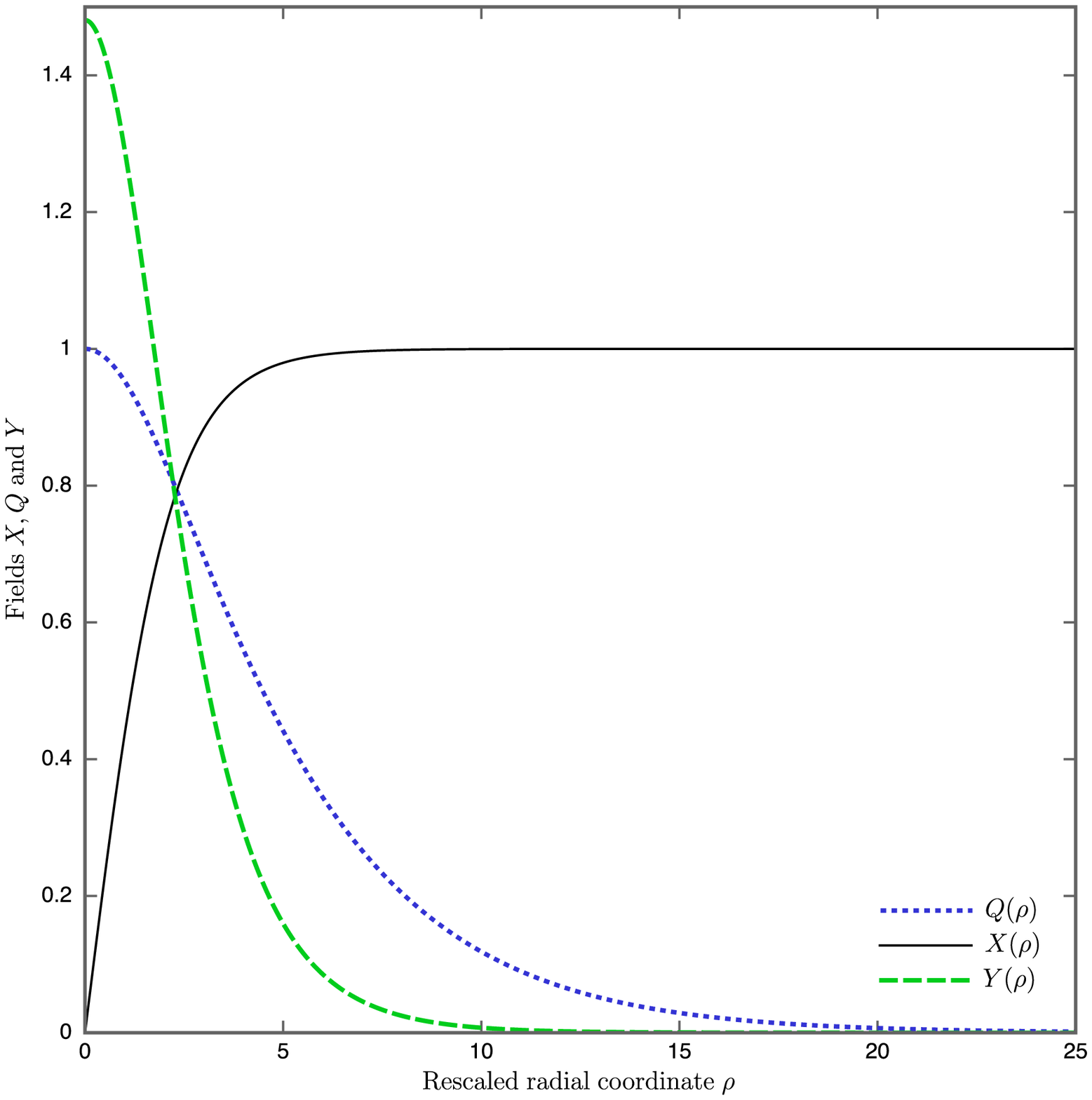}\hskip1cm
\hskip-5mm\includegraphics[scale=.47]{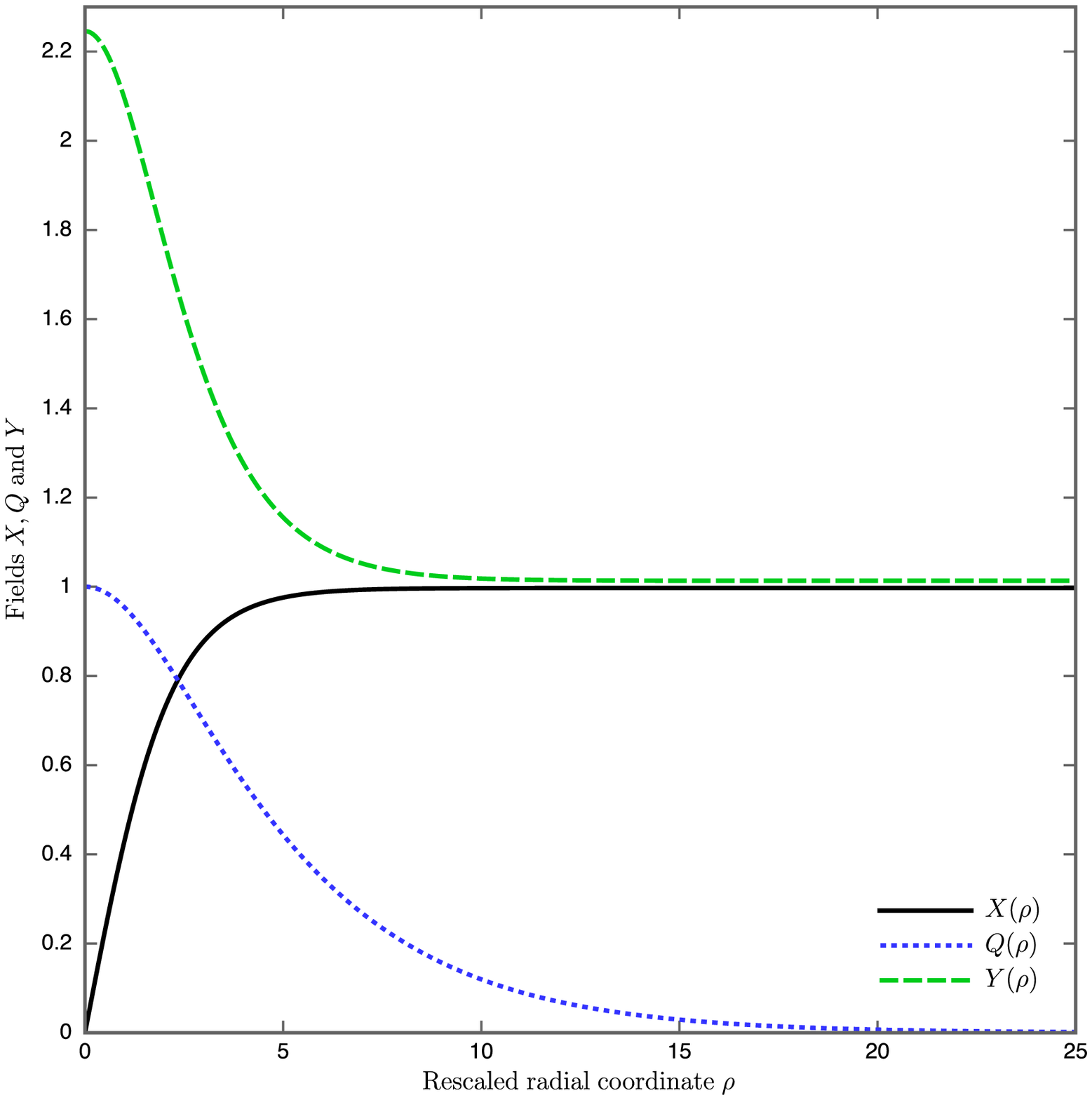}
\caption{Typical numerical solution of the system (\ref{varphi}),
  (\ref{Q}) and (\ref{profile}) with constant phases ($\alpha$,
  $\beta$ and $\gamma$ constant) for the dimensionless fields
  $X(\rho)=\varphi/\eta$, $Y(\rho)=\sigma \sqrt{\lambda_\sigma} /
  m_\sigma$, and $Q$ as function of the rescaled distance to the
  string core $\rho = \sqrt{\lambda_\phi} \eta r$ for parameters fixed
  to $\tilde q^2=0.1$, $\alpha_1 = 3.37\times 10^{-3}$, $\alpha_2 =
  2.63\times 10^{-3}$, and $\alpha_3 = 5.26\times 10^{-4}$; the
  relevant free parameters are defined in a way reminiscent of
  Ref.~\cite{neutral}, i.e. $\tilde q^2=2q^2/\lambda_\phi$, $\alpha_1
  = m_\sigma^2/(2\lambda_\sigma \eta^2)$, $\alpha_2 = f m_\sigma^2 /
  (2\lambda_\phi \lambda_\sigma \eta^2)$ and $\alpha_3 =m_\sigma^4 /
  (2\lambda_\phi \lambda_\sigma \eta^4)$ (the $\alpha_i$'s have of
  course nothing to do with the angle $\alpha$ introduced before). The
  solutions are calculated by means of Successive Over Relaxation
  \cite{sor} for both cases for which either the SU(2) field does not
  condense in vacuum, i.e. for the plus sign in front of the massive 
  term in Eq.~(\ref{PotSig}) (left panel) and that
  for which the SU(2) field does condense in vacuum, i.e. for
  the minus sign (right panel). The fact that the three curves for
  the Higgs field [$X(\rho)$, full line], the U(1) gauge field
  [$Q(\rho)$, dashed line] and the SU(2) scalar condensate
  [$\sigma(\rho)$, dotted line] seem to cross at a single point for
  the non condensing case of the left panel is purely coincidental and
  merely due to the specific choice of the parameters. The
  normalization of $Y$ with respect to that of $\sigma$ implies that
  in the large distance limit $\rho\to\infty$, one has $Y\to\frac12
  (1-\varepsilon)$.
  \label{Bckd}}
\end{figure*}

In fact, they can still be further simplified. Indeed, let us now
expand the vector components in such a way as to implement its
normalization, i.e. by projecting these components on the sphere on
which it evolves in terms of angular variables $\beta(t,r,z,\theta)$
and $\gamma(t,r,z,\theta)$. This gives
\begin{eqnarray}
n^1 &=& \sin\beta\sin\gamma,\cr
n^2 &=& \sin\beta \cos\gamma,\cr
n^3 &=& \cos\beta,
\label{betagamma}
\end{eqnarray}
and therefore 
\begin{equation}
n_a \Dalembert n^a = -\left(\partial\beta\right)^2 -\sin^2\beta
\left(\partial\gamma\right)^2,
\label{naBoxnabetagamma}
\end{equation}
which shows that Eq.~(\ref{alphadyn}) is indeed a dynamical
equation for the variable $\alpha$ only. Using the expansion
(\ref{betagamma}), one can transform Eq.~(\ref{tauasimple}) into
\begin{equation}
\Dalembert\beta + 2 \left( \frac{\partial\sigma}{\sigma} +
 \frac{\partial\alpha}{\tan\alpha} \right) \cdot \partial \beta =
\cos\beta\sin\beta\left(\partial\gamma \right)^2,
\label{beta}
\end{equation}
and
\begin{equation}
\Dalembert\gamma + 2 \left( \frac{\partial\sigma}{\sigma} +
 \frac{\partial\alpha}{\tan\alpha}+ \frac{\partial\beta}{\tan\beta}
\right) \partial\gamma =0,
\label{gamma}
\end{equation}
that completes a new set of dynamical equations, namely
Eqs.~(\ref{alphadyn}), (\ref{alphadyn}), (\ref{beta})
and~(\ref{gamma}), for the 4 independent functions $\sigma$, $\alpha$,
$\beta$ and $\gamma$. A particular solution for constant angles and
gradients (lowest energy state) is exemplified in Fig.~\ref{Bckd} for
the cases for which SU$(2)$ is unbroken or broken far from the string,
derived using typical values for the parameters.

\section{abelian cases}
\label{abeliansol}

Since the group SU$(2)$ contains invariant U$(1)$s, it can be used,
restricting to special cases, to recover the abelian Witten
model~\cite{Witten:1984eb} as well as the biabelian
case~\cite{LPXcoupled}. The purpose of this section is precisely to
establish the correspondences.

\subsection{Witten abelian model}
\label{ab}

The form (\ref{Sigma}) for the scalar doublet can be rewritten in
terms of the angles $\alpha$, $\beta$ and $\gamma$ as
\begin{equation}
\bm{\Sigma} = \frac{\sigma}{\sqrt{2}} \left( \begin{array}{c}
\ex^{i\gamma}\sin\alpha\sin\beta\cr \cos\alpha-i\sin\alpha\cos \beta
\end{array}\right),
\label{Sigabgam}
\end{equation}
{}from which one would like to single out a phase representing the
U$(1)$ situation. In other words, one wants to identify real functions
$\psi$, $f$ and $g$ such
that 
\begin{equation}
\bm{\Sigma} = \ex^{i\psi}  \left( \begin{array}{c}
f \cr g
\end{array}\right).
\label{SigTot}
\end{equation}
Through identification of (\ref{SigTot}) with (\ref{Sigabgam}),
one can easily convince oneself that there are only two possibilities,
namely
\begin{equation}
\alpha=\beta=\frac{\pi}{2}, \ \ \ \ \psi = \gamma, \ \ \ \
f=\frac{\sigma}{\sqrt{2}}, \ \ \ \ g=0, 
\label{ab1}
\end{equation}
and
\begin{equation}
\psi =- \alpha, \ \ \ \beta=0,\ \ \ \ \gamma\in\setR, 
\ \ \ f=0, \ \ \ \ g=\frac{\sigma}{\sqrt{2}}.
\label{ab2}
\end{equation}
The first case, Eq.~(\ref{ab1}), leads to $n_a\Dalembert n^a =
-\left(\partial\psi\right)^2$, and the field equations become
\begin{equation}
\Delta \sigma - \left(\partial\psi\right)^2 \sigma = \frac12
\frac{\partial V}{\partial \sigma},
\label{sigab1}
\end{equation}
and 
\begin{equation}
\Dalembert \psi + \frac{2}{\sigma}\frac{\dd\sigma}{\dd r} 
\partial_r \psi =0.
\label{psiab1}
\end{equation}
In the abelian case, the phase does not depend on the radial distance
and, hence, the last equation simply becomes $\Dalembert \psi=0$. This
relation, together with Eq.~(\ref{sigab1}), are exactly the equations
of motion in the abelian case
\cite{Witten:1984eb,formal3,carterPLB89,formal2,formal4,neutral,
  enon0,NoSpring}. The fact that we recover them from the most general
framework discussed here is a consistency check of
Eqs.~(\ref{alphadyn}), (\ref{alphadyn}), (\ref{beta})
and~(\ref{gamma}). In the same manner, one can also check that the
ansatz~(\ref{ab2}) also leads to the abelian equations of motion.

\par

At this point, a clarification concerning the abelian situation is
useful. With the set of equations above, one in principle assume the
phase to vary only along the worldsheet directions, i.e.,
$\psi=\psi(z,t)$, see above. However, this is not merely an
assumption, but rather a fact that can be demonstrated through
separation of variables: since the scalar field amplitude $\sigma$
depends only on the radial distance $r$, setting $\psi = R(r) +
T(\theta) + W(z,t)$, Eq.~(\ref{sigab1}) tells us
that
\begin{equation} 
\left(\partial\psi\right)^2 = \left(\frac{\dd R}{\dd
    r}\right)^2 +\frac{1}{r^2}\left(\frac{\dd T}{\dd \theta}\right)^2
+\left(\partial_z W\right)^2 -\left(\partial_t W\right)^2,
\end{equation}
is a yet unknown function of $r$ only, which we write temporarily as
$f(r)$. This implies that $T=T_0+p\theta$, and hence
\begin{equation}
\label{profilephaseabelian}
\left(\frac{\dd R}{\dd r}\right)^2 +\frac{p^2}{r^2} -f(r) =
-\left(\partial_z W\right)^2 +\left(\partial_t W\right)^2 \equiv -w,
\end{equation}
where $w$ is a separation constant, to be later identified with the
state parameter of the abelian current-carrying cosmic string. The
equation $\left(\partial_z W\right)^2 -\left(\partial_t W\right)^2 =w$
can also be solved trough separation of variables. Indeed, writing
$W(z,t)$ as the sum of a function of $z$ and of a function of $t$, one
can show that these two functions are in fact linear in $z$ and $t$
respectively.

\par

Similarly separating variables in Eq.~(\ref{psiab1}) then leads to
\begin{equation}
\frac{\dd^2 R}{\dd r^2} + \left(\frac{1}{r} + \frac{2}{\sigma}\right)
 \frac{\dd \sigma}{\dd r} \frac{\dd R}{\dd r} = \partial^2_t W
 - \partial^2_z W=0,
\label{R2}
\end{equation}
since we have just seen that $W$ is the sum of two linear functions
(and, therefore, its second order derivatives vanish). This can be
integrated to yield
\begin{equation}
\frac{\dd R}{\dd r} = \frac{A}{r\sigma^2},
\label{Rp}
\end{equation}
where $A$ is a constant. If we insert this expression into
Eq.~(\ref{profilephaseabelian}), this leads to an explicit expression
for the function $f(r)$, namely
\begin{equation}
f(r)=w+\frac{p^2}{r^2}+\frac{A^2}{r^2\sigma ^4}=
\left(\partial \psi\right)^2.
\end{equation}
This function must be plugged back into Eq.~(\ref{sigab1}) in order to
obtain the full profile. Since $\dd R/\dd r \propto r^{-1}
\sigma^{-2}$, there is no way to obtain a regular solution for
$\sigma$ unless the constants $p$ and $A$ are made to vanish,
i.e. unless $R(r)$ is in fact a constant. One recovers the possibility
to concentrate on pure worldsheet phase excitations, and the dynamics
of the worldsheet merely depends on the phase gradients, the state
parameter. It is important to notice at this stage that the second
derivatives of the phase do contribute neither at the level of the
field equations, nor at that of the stress tensor: this is why one
usually disregards them and sets, without loss of generality, the phase
as $\psi = k z -\omega t$, with the state parameter being
$w=k^2-\omega^2$.

\subsection{The biabelian case}
\label{biabelian}

One step further in the direction of a full nonabelian situation is
that of two abelian currents, dubbed the biabelian current-carrying
string, as was in particular studied in Ref.~\cite{LPXcoupled}. In
this case, one identifies a U$(1)\times$U$(1)$ piece in SU$(2)$
through the requirement
\begin{equation}
\bm{\Sigma}
\equiv \left( \begin{array}{c}
\Sigma_1\cr \Sigma_2 \end{array}\right) = \frac{1}{\sqrt{2}}
\left( \begin{array}{c} 
\sigma_1 \ex^{i\psi_1}\cr \sigma_2 \ex^{i\psi_2}
\end{array}\right).
\label{biab}
\end{equation}
There is no direct identification that can be done here for which the
phases, contrary to the actual biabelian one, would depend only on the
worldsheet coordinates: this is due to the fact that SU$(2)$ is
topologically equivalent to a 3-sphere, whereas the U$(1)\times$U$(1)$
we consider consists in two independent circles at the surface of this
3-sphere. As the phases vary, in principle independently, around the
circles, they cannot describe an actual trajectory along the 3-sphere,
hence the problem.

\par

Thus, there cannot be a simply defined global solution in this
case. It turns out that, in order to recover the actual
U$(1)\times$U$(1)$, one must apply a trick, which we shall also use
afterwards in the full nonabelian case. It consists in first
identifying the phases as
\begin{equation}
\psi_1=\gamma, \ \ \ \ \psi_2 = -\tan^{-1} \left(\cos\beta \tan\alpha \right),
\label{biabphases}
\end{equation}
so that the amplitudes are given by
\begin{equation}
\sigma_1^2 = \sigma^2 \sin^2\alpha \sin^2\beta, \ \ \ \ \sigma_2^2 =
\sigma^2 \left( \cos^2\alpha + \sin^2\alpha \cos^2\beta\right).
\label{biabampl}
\end{equation}
We immediately see where the problem originates, because in principle
one expects the phases to depend on $z$ and $t$, while the amplitude
should be functions of the string radial distance $r$. But in the case
of Eqs.~(\ref{biabphases}) and (\ref{biabampl}), one phase, namely
$\psi_1=\gamma$, enters independently of the rest and can therefore
safely be assumed to vary along $z$ and $t$, but the second phase and
the amplitudes involve the \emph{same} functions in a essentially
nonlinear way.

\par

The way to recover the previous case is to assume an ultralocal
hypothesis, which consists in saying that the fields are to be
evaluated at only one point of the worldsheet, which we set, for
simplicity, to be at $z=t=0$, while we keep the gradients at this very
point. This means in practice that we consider the angles as functions
of the radial distance and set their gradients along the string to
\begin{equation}
\partial_z\alpha\to k_\alpha \ \ \
\hbox{and} \ \ \ \partial_t\alpha\to -\omega_\alpha,
\label{ultra}
\end{equation}
and similar relations for $\beta$ and $\gamma$.

\par

The kinetic term $K=|\partial\Sigma_1|^2 +|\partial\Sigma_2|^2$ in the
action then becomes
\begin{equation}
K= \frac12 \left\{ \sigma'^2
 +\sigma^2 \left[ \alpha'^2 + w_\alpha + \sin^2\alpha\left(
     w_\beta+w_\gamma\right)\right]\right\},
\label{Kbiab}
\end{equation}
where a prime denotes a derivative w.r.t. $r$ and we have set
$w_i\equiv k_i^2 - \omega_i^2$ for each angle $i\in\{\alpha,
\beta,\gamma\}$. Taking into account the identifications
(\ref{biabphases}) and (\ref{biabampl}), we see that provided we
write $w_1 = w_\alpha+w_\beta +w_\gamma$ and $w_2=w_\alpha$, it takes
the canonical form for two scalar current-carriers, namely
\begin{equation}
K=\frac12 \left( \sigma_1'^2 + \sigma'^2 + w_1 \sigma_1^2 + w_2
 \sigma_2^2 \right). 
\label{twofields}
\end{equation}

\par

In the final section, devoted to the stress energy tensor of the
string, we shall discuss the conditions on the parameters, for it can
easily be seen right away that at this stage, the model contains 6
independent parameters (the phase gradients), whereas we know that the
actual U$(1)\times$U$(1)$ case can be fully described with only 3,
which are the worldsheet Lorentz invariants that can be built out of
the two phase gradients. The fact that the string stress tensor can
only depend on Lorentz invariant quantities must be implemented by
hand at this stage, and it gives precisely the exact values for the
eigenvalues that are the energy per unit length and the tension. The
ultralocal procedure described below is thus validated in this case.

\section{The nonabelian part}
\label{nonabelsol}

Let us first build on the second solution of Sec.~\ref{ab}
[Eq.~(\ref{ab2})] and assume that $\alpha$
depends on the external coordinates and is function of $z$ and $t$
only. We will show that this implies that $\beta $ and $\gamma $ also
depend only on $z$ and $t$; this would be the most natural
generalization of the Witten model for which the phase excitation only
move along the worldsheet. However, we find that there is only one
such globally defined solution, containing three chiral propagation
modes. Let us see how this happens.

\subsection{An Exact Solution: the trichiral Case}
\label{subthreechiral}

Let us start with seeking solutions for the angle $\alpha$. Looking at
Eq.~(\ref{Sigma}), one notices that the term $\cos \alpha $ represents
a natural abelian part of the solution since only this term remains if
one requires $n^a=0$. In other words, $\alpha$ again identifies a
subgroup U(1) of the original SU(2) along which the condensate behaves
as a usual abelian current-carrying cosmic string. In this situation,
one also recovers the previously discussed abelian solution. As a
consequence, it seems natural to assume that $\alpha $ is a function
of $z$ and $t$, so that
\begin{equation}
\partial\alpha\cdot \partial\sigma = 0.
\label{daldsig}
\end{equation}
Moreover, as $\sigma$ depends only on $r$, it is immediately clear
from Eq.~(\ref{Id}) that
\begin{equation}
\left(\partial\alpha\right)^2 + \tan\alpha
\Dalembert \alpha = w,
\label{w}
\end{equation}
where $w$ is a constant, again to be later identified with the state
parameter of the abelian current-carrying string. Plugging the
relation (\ref{w}) back into Eq.~(\ref{tauana}) now gives the
constraint
\begin{equation}
n_a \Dalembert n^a = -\frac{2\Dalembert\alpha}{\sin 2\alpha} =
\frac{\left(\partial\alpha\right)^2 - w}{\sin^2 \alpha}.
\label{naBoxna}
\end{equation}
Eq.~(\ref{w}) can be solved setting $u=\cos\alpha$ as it then
transforms into the linear Klein-Gordon equation
\begin{equation}
\left( \Dalembert -w\right) u = 0,
\label{linu}
\end{equation}
whose general solution is easily obtained. It reads
\begin{eqnarray}
u &=& \cos \left(\omega t-kz-\alpha_0\right)
+\int \Bigl[ s_+(E) \ex^{i\left(Et+\sqrt{E^2-w} z\right)} 
\nonumber \\
&+&
s_-(E) \ex^{ i\left(Et-\sqrt{E^2-w} z\right)} \Bigr] \dd E ,
\label{solu}
\end{eqnarray}
with $s_\pm(E)$ two arbitrary (unknown) functions of $E$ and $w\equiv
k^2-\omega^2$. This general solution is made of two pieces. The first
one, 
\begin{equation}
\label{alpha}
\alpha = \alpha_0 + kz-\omega t, 
\end{equation}
is the exact equivalent of the U(1) conducting string phase. Note that
this was to be expected since, as mentioned above, $\alpha$ picks a
special U(1) direction of the original SU(2) \cite{BCpriv}. At this
point however, it is worth mentioning that contrary to the U(1) case,
there is no simple way to cancel out the constant $\alpha_0$ appearing
: since a simple SU(2) transformation can never be expressed as a
shift in $\alpha$, one cannot simply set $\alpha_0\to 0$, so that this
quantity is actually endowed with a physical (measurable) meaning. The
second part of the solution represents massive particles moving along
the worldsheet when one considers usually normalized distribution
functions $s_\pm$. We are however interested in collective modes along
the string, and therefore restrict attention to the special case for
which $s_\pm =0$. Let us also notice that, if $w=0$, then $u$ becomes
an arbitrary function of $t+z$ and $t-z$. Inserting this solution back
into Eq.~(\ref{w}), we see that $\alpha $ becomes an arbitrary
function of $t+z$ or $t-z$,
\begin{equation}
\alpha_{\mathrm{chiral}} = \alpha\left( t+\varepsilon z \right),
\ \ \ \hbox{with} \ \ \ \varepsilon = \pm 1.
\label{alphachiral}
\end{equation}
To summarize, we have two possible situations: either $w\neq 0$ and one
must consider the solution~(\ref{alpha}) or $w=0$ and one must work
with the chiral solution given by~(\ref{alphachiral}).

\par

Finally, we notice that, for the two above mentioned cases, one has
$\Dalembert \alpha=0$ which in turn, thanks to Eq.~(\ref{naBoxna}),
means
\begin{equation}
\label{naBoxnazero}
n_a \Dalembert n^a=0 \, .
\end{equation}

\par

We then look for a nontrivial solution for the vector $n^a$ whose
dynamics is given by Eq.~(\ref{tauasimple}). Once one takes into
account that $\alpha $ is a function on $z$ and $t$ only, see
Eqs.~(\ref{alpha}) or~(\ref{alphachiral}), this relation reduces to
\begin{equation}
\Dalembert n^a + 2 \frac{\dd\ln\sigma}{\dd r} \partial_r n^a +
\frac{2}{\tan\alpha}\left[\left(\partial _z\alpha\right) \partial_z
-\left(\partial _t\alpha \right)\partial_t\right] n^a =0.
\label{naw}
\end{equation}
Therefore, one must solve this equation together with the
constraint~(\ref{naBoxnazero}), $n_a \Dalembert n^a = 0$.

\par

We first rewrite Eq.~(\ref{naw}) as dynamical equations for the
worldsheet functions $\beta$ and $\gamma$. We find
\begin{widetext}
\begin{equation}
\frac{\partial^2\beta}{\partial
r^2}+\left(\frac{1}{r}+2\frac{1}{\sigma} \frac{\dd\sigma}{\dd r}
\right) \frac{\partial\beta}{\partial r} +\frac{1}{r^2}
\frac{\partial^2\beta}{\partial\theta^2} +
\frac{\partial^2\beta}{\partial z^2} - \frac{\partial^2\beta}{\partial
t^2} + \frac{2}{\tan \alpha} \left[\left(\partial _z\alpha \right)
\frac{\partial\beta}{\partial
  z}-\left(\partial _t\alpha\right)
\frac{\partial\beta}{\partial t}\right] = 0,
\label{mvtbeta}
\end{equation}
and
\begin{equation}
\frac{\partial^2\gamma}{\partial
r^2}+\left(\frac{1}{r}+2\frac{1}{\sigma} \frac{\dd\sigma}{\dd r}
\right) \frac{\partial\gamma}{\partial r} +\frac{1}{r^2}
\frac{\partial^2\gamma}{\partial\theta^2} +
\frac{\partial^2\gamma}{\partial z^2} - \frac{\partial^2\gamma}{\partial
t^2} + \frac{2}{\tan \alpha} \left[\left(\partial _z\alpha\right)
\frac{\partial\gamma}{\partial
  z}-\left(\partial _t\alpha\right)
\frac{\partial\gamma}{\partial t}\right] = 0,
\label{mvtgamma}
\end{equation}
\end{widetext}
showing that $\beta$ and $\gamma$ are subject to the same dynamics, so
that their potentially different behaviors merely rely on their
initial conditions. One the other hand, the
constraint~(\ref{naBoxnazero}) reads
\begin{equation}
\left(\partial\beta\right)^2 + \sin^2\beta\left( \partial\gamma\right)^2 =0,
\label{naBoxna0}
\end{equation}
showing that, in the four dimensional embedding spacetime, the phase
gradients $\partial_\mu\beta$ and $\partial_\mu\gamma$ are
lightlike. However, this is not the end of the discussion, for the
fields $\beta$ and $\gamma$ actually live in the embedding
four-dimensional space-time. They could therefore vary, in a
lightlike way, in all directions around the vortex, and after
integration over the transverse degrees of freedom, leave the
appearance of a spacelike or timelike variation. This, in fact, is to
be expected on general geometrical considerations~\cite{BCIII,BCIV},
leading to many equation of state parameters. We shall see below that
it is not what happens in the case at hand. Concretely,
Eqs.~(\ref{naBoxna0}) amount to
\begin{eqnarray}
\label{consbeta}
\left(\frac{\partial\beta}{\partial r}\right)^2+\frac{1}{r^2}
\left(\frac{\partial\beta}{\partial\theta}\right)^2 +
\left(\frac{\partial\beta}{\partial z}\right)^2 
- \left(\frac{\partial\beta}{\partial t}\right)^2=0, 
\\
\left(\frac{\partial\gamma}{\partial r}\right)^2+\frac{1}{r^2}
\left(\frac{\partial\gamma }{\partial\theta}\right)^2 +
\left(\frac{\partial\gamma }{\partial z}\right)^2 
- \left(\frac{\partial\gamma }{\partial t}\right)^2=0.
\label{consgamma}
\end{eqnarray}
These equations have the form of two gravitational Hamilton-Jacobi
equations, that is to say $g^{ik}(\partial S/\partial x^i)(\partial
S/\partial x^k)=0$. Consequently, they can be explicitly solved by
means of separation of variables. Setting $\beta = R_\beta(r)
T_\beta(\theta) b_\beta(z,t)$ and $\gamma = R_\gamma(r)
T_\gamma(\theta) b_\gamma(z,t)$, the complete system of equations
reads
\begin{eqnarray}
\label{consT}
& &\left(\frac{{\rm d}T_i}{\dd \theta}\right)^2-\nu_i^2T_i^2=0, \\
\label{consR}
& &\left(\frac{{\rm d}R_i}{\dd r}\right)^2-\frac{1}{r_i^2}
\left(1+\nu_i^2\frac{r_i^2}{r^2}\right)R_i^2=0,\\
\label{consb}
& & \left(\frac{\partial b_i}{\partial z}\right)^2
-\left(\frac{\partial b_i}{\partial t}\right)^2
+\frac{b_i^2}{r_i^2}=0\, ,
\end{eqnarray}
for both $i=\beta$ and $\gamma$, where $r_i$ and $\nu_i$ are
separation constants. Of course, the solution must also satisfy the
dynamical equations~(\ref{mvtbeta})
and~(\ref{mvtgamma}). Straightforward manipulations show that this
amounts to
\begin{eqnarray}
\label{dynT}
& &\frac{{\rm d}^2T_i}{\dd \theta^2}-\mu_i^2T_i=0, \\
\label{dynR}
& &\frac{{\rm d}^2R_i}{\dd r^2}+\left(\frac{1}{r}+\frac{2}{\sigma}
\frac{\dd \sigma}{\dd r}\right)\frac{\dd R_i}{\dd r}-
\left(w_i+\frac{\mu_i^2}{r^2}\right)R_i=0,\\
\label{dynb}
& & \frac{\partial ^2b_i}{\partial z^2}
-\frac{\partial ^2b_i}{\partial t^2}+\frac{2}{\tan \alpha }
\left(\partial _z\alpha
\frac{\partial b_i}{\partial z}-\partial _t\alpha
\frac{\partial b_i}{\partial t}
\right)+w_ib_i=0,\cr & & 
%\right]\nonumber \\ & & +w_ib_i=0,
\end{eqnarray}
where $w_i$ and $\mu_i$ are two new constants of separation. The main
question is now whether the solutions obtained from the dynamical
equation are compatible with the ones derived from the constraint.

\par

The two equations~(\ref{consR}) and~(\ref{dynR}) controlling the
behavior of $R_i$ can only be compatible if the function $R_i$ is a
constant since one of this equation, Eq.~(\ref{dynR}), contains
$\sigma $ while the other, Eq.~(\ref{consR}) does not. This
immediately implies $\nu_i=1/r_i=\mu_i=w_i=0$ and we are left with
\begin{equation}
\left(\frac{\partial b_i}{\partial z}\right)^2
-\left(\frac{\partial b_i}{\partial t}\right)^2=0,
\end{equation}
and
\begin{equation}
\frac{\partial ^2b_i}{\partial z^2}
-\frac{\partial ^2b_i}{\partial t^2}+\frac{2}{\tan \alpha }
\left[\left(\partial _z\alpha\right)
\frac{\partial b_i}{\partial z}-\left(\partial _t\alpha\right)
\frac{\partial b_i}{\partial t}
\right]=0.
\end{equation}
Of course, one possibility is to take $b_i$ as constant. However, this
means that the vector $n^a$ is fixed and this just corresponds to the
abelian case. In fact the general solution of the first equation above
is $b_i=b_i(t+\varepsilon_iz)$, with $\varepsilon_i=\pm 1$. Inserting
this solution into the second relation, one obtains
\begin{equation}
\varepsilon_i\left(\partial _z\alpha\right)
-\left(\partial _t\alpha\right)=0\, .
\end{equation}
If $\alpha $ is given by Eq.~(\ref{alpha}), then the above equation
becomes $\omega =-\varepsilon_ik$ which implies $w=0$. But, if $w=0$,
then one must consider the chiral solution~(\ref{alphachiral}). In
this case, the dynamical solution reduces to $\varepsilon
\varepsilon_i=1$. This means that one also obtains chiral solutions
for these angles, namely
\begin{equation}
\beta_{\mathrm{chiral}} = \beta\left( t+\varepsilon z \right),
\ \ \ \hbox{with} \ \ \ \varepsilon = \pm 1,
\label{betachiral}
\end{equation}
and
\begin{equation}
\gamma_{\mathrm{chiral}} = \gamma\left( t+\varepsilon z \right),
\ \ \ \hbox{with} \ \ \ \varepsilon = \pm 1,
\label{gammachiral}
\end{equation}
We see that this solution contains three chiral-like functions, hence its
name. It is of course very important to notice that the relative sign
in the argument of $\alpha$, $\beta $ and $\gamma $ needs to be the
same for these three functions. This implies that all the angles must
propagate in the same direction, i.e. the string currents consist in
right or left movers only. The situation is thus the same as that
first discussed in Ref.~\cite{CP99}, but with three
independent copies of the currents and the additional constraint that
they all move in the same direction.

\par

Constructing a surface action over the wordsheet (with coordinates
$\xi_i)$ 
\begin{equation}
\mathcal{S} = \int \dd^2 \xi \sqrt{-h}\mathcal{L}^{(2)}(\xi_i),
\label{S2}
\end{equation}
for such a trichiral string is a
straightforward generalization of \cite{CP99}: if one assumes
a two dimensional Lagrangian of the form
\begin{equation}
\mathcal{L}^{(2)} = - m^2 -\frac12 \mathcal{M}^{\si{AB}}
h^{ij} \partial_i \psi_{\si{A}}\partial_j \psi_{\si{B}},
\label{L2chiral}
\end{equation}
where $m$ is a constant describing the Nambu-Goto string background
and $\mathcal{M}^{\si{AB}}$ is a matrix Lagrange multiplier with no
kinematic term in the action, $h^{ij}$ is the worldsheet induced
metric and the $\psi_{\si{A}}$ stand for our angular functions
$\alpha$, $\beta$ and $\gamma$. Varying with respect to this matrix
immediately provides the null conditions for all the fields, namely
\begin{equation}
h^{ij} \partial_i \psi_{\si{A}}\partial_j \psi_{\si{B}} = 0,
\label{chir}
\end{equation}
showing that not only all the fields are lightlike, but also, if the
matrix $\mathcal{M}$ is non diagonal, that all the solutions do move
in the same direction, i.e. that they are all either right or left movers.

\subsection{A No-Go Theorem for Exact Separable Solutions}
\label{seunoexact}

In fact, one can show that the trichiral solution is the only exact
separable solution. Indeed, Eq.~(\ref{consR}) can be easily
solved. Its solution reads
\begin{equation}
\frac{R_i}{R_i^0}=\exp\left(\pm \frac{r}{r_i}
\sqrt{1+\nu_i^2\frac{r_i^2}{r^2}}\right)
\left(\nu_i\frac{r_i}{r}+\sqrt{1+\nu_i^2\frac{r_i^2}{r^2}}\right)^{\mp\nu_i},
\end{equation}
where $R_i^0$ is an integration constant. However, inserting this
expression into Eq.~(\ref{dynR}) shows that it is solution only if
$R_i$ is a constant. This is of course due to the presence of the term
$(\dd \sigma/\dd r)/\sigma $ which cannot be canceled by any other
term. But if $R_i$ is a constant, then $\nu_i=0$ which in turn implies
that $T$ is also a constant. In other words, we are back to the
trichiral solution of the previous section.

\par

This shows that there is no other exact and separable
solution. Although this, of course, does not, in principle, prevent
the existence of solutions which do not obey separation of variables,
there exists a general argument, due to Carter \cite{BCIII,BCIV},
showing that one should not expect a global solution to exist. The
argument relies on the fact that the generators of the currents form a
manifold whose curvature is non zero, while the cylindrically
symmetric string configuration assumes vanishing extrinsic and
intrinsic curvatures, thus leading to an incompatibility.

\section{Ultralocal crooked string}

The SU$(2)$ condensate does not have any regular nontrivial solution
expect for the trichiral: does this mean that only abelian or
chiral-like current-carrying cosmic strings can be formed?

\par

The answer to this question involves two different perspectives. First,
one must remember that when the current builds up along the string, it
does so through a random process through which phases take
uncorrelated values on distances larger than the correlation
length. There is therefore no reason to assume the current would be,
all along the worldsheet, always following one particular U$(1)$
direction. Moreover, all the above discussion heavily relies on a
straight and static string whose fundamental tensor is merely the two
dimensional Minkowski metric. The string manifold, therefore, is
described as flat, and this is the cause for the discrepancy: SU$(2)$
having a nonvanishing curvature, it is normal that it cannot be
projected onto the string worldsheet, so only a flat subspace of it,
the U$(1)$ we identified, remains once this operation is performed.

\par

The way to reconcile both perspectives is by considering an actual
string, which, as simulations reveal, is in fact crooked, and
definitely not flat. Locally, one can always approximate the string by
a straight line, and assume cylindrical symmetry. However, this is
only a rough approximation which, although valid in the abelian case,
is severely limited in the nonabelian case. In order to take into
account the possible variations of the phases without having a
solution satisfying the requirement of cylindrical symmetry, we
introduce a so-called ultralocal approximation, by which we restrict
attention to one particular point on the worldsheet, which we take for
simplicity (and without lack of generality), to be at $z=t=0$, but
keep the phase gradients along the worldsheet as parameters. This
procedure, applied to the abelian and biabelian cases, gives the
correct result.

\par

In practice, the ultralocal approximation for the crooked nonabelian
current-carrying cosmic string consists in assuming the phases to
depend on the radial distances, while their gradients are numbers. In
other words, we set
\begin{equation}
\alpha \to \alpha(r) + k_\alpha z - \omega_\alpha t + \frac12 \left(
\alpha^0_{,zz} z^2 + \alpha^0_{,tt} t^2\right) + \cdots
\label{alcroo}
\end{equation}
(and similar expressions for $\beta$ and $\gamma$) and let $z,t\to 0$
in the final expressions we obtain. Note that this procedure only
applies in the very final equations, and for instance it is not
possible to apply it for the action itself, as the field equations
derived from the approximated action would not be equivalent to the
approximated field equations derived from the exact action, lacking in
particular the squared gradients and second derivatives with respect
to the worldsheet coordinates.

\par

Using the approach described above, it is straightforward to derive the
equations of motion obeyed by the three angles $\alpha $, $\beta $ and
$\gamma$. Since we are interested in the minimal energy configuration,
we ignore a possible $\theta $ dependence. As a consequence, only
equations controlling the profiles of the functions $\alpha (r)$,
$\beta (r)$ and $\gamma (r)$ remain. They read
\begin{widetext}
\begin{eqnarray}
& & \frac{{\rm d}^2\alpha }{{\rm d}r^2}+\frac{1}{r}
\frac{{\rm d}\alpha }{{\rm d}r}
+2\frac{{\rm d}\sigma }{{\rm d}r}\frac{{\rm d}\alpha }{{\rm d}r}
+\alpha^0_{,zz}-\alpha^0_{,tt}
-\sin \alpha \cos \alpha \left[\left(\frac{{\rm d}\beta }{{\rm d}r}\right)^2
+k_{\beta }^2-\omega _\beta ^2\right]
-\sin \alpha\cos \alpha \sin ^2\beta
\left[\left(\frac{{\rm d}\gamma }{{\rm d}r}\right)^2
+k_{\gamma }^2-\omega _\gamma ^2\right]=0, 
\\
& & \frac{{\rm d}^2\beta }{{\rm d}r^2}+\frac{1}{r}\frac{{\rm d}\beta }{{\rm d}r}
+2\frac{{\rm d}\sigma }{{\rm d}r}\frac{{\rm d}\beta }{{\rm d}r}
+\beta^0_{,zz}-\beta^0_{,tt}
+\frac{2}{\tan \alpha}\left(\frac{{\rm d}\alpha}{{\rm d}r}
\frac{{\rm d}\beta }{{\rm d}r}+k_{\alpha}k_{\beta}-\omega _{\alpha}\omega _{\beta}
\right)-\sin \beta \cos \beta
\left[\left(\frac{{\rm d}\gamma}{{\rm d}r}\right)^2
+k_{\gamma }^2-\omega _\gamma ^2\right]=0,
\\
& & \frac{{\rm d}^2\gamma}{{\rm d}r^2}+\frac{1}{r}
\frac{{\rm d}\gamma }{{\rm d}r}
+2\frac{{\rm d}\sigma }{{\rm d}r}\frac{{\rm d}\gamma }{{\rm d}r}
+\gamma^0_{,zz}-\gamma^0_{,tt}
+\frac{2}{\tan \alpha}\left(\frac{{\rm d}\alpha}{{\rm d}r}
\frac{{\rm d}\gamma }{{\rm d}r}+k_{\alpha}k_{\gamma}-\omega _{\alpha}
\omega _{\gamma}\right)
+\frac{2}{\tan \beta}\left(\frac{{\rm d}\beta}{{\rm d}r}
\frac{{\rm d}\gamma }{{\rm d}r}+k_{\beta}k_{\gamma}-\omega _{\beta}
\omega _{\gamma}\right)=0.
\end{eqnarray}
\end{widetext}
As expected, the profiles depend on the six parameters $k_i$ and
$\omega _i$. However, and this a new feature of the nonabelian case,
there is also an additional dependence in the second order derivatives
which introduces three new Lorentz invariant parameters, namely
$\alpha^0_{,zz}-\alpha^0_{,tt}$, $\beta^0_{,zz}-\beta^0_{,tt}$ and
$\gamma^0_{,zz}-\gamma^0_{,tt}$.

\par

The shape of the profiles will be very similar to what one encounters
in the abelian case as a simple study of the behavior of the above
equations in the limit $r\rightarrow 0$ and $r\rightarrow +\infty$
reveals. The precise form of the profiles does not bring much insight
into the problem at hand and, therefore, we now turn to the
calculation of the stress-energy tensor.

\section{Wordsheet stress energy tensor}
\label{secTmn}

Our aim is to describe the string worldsheet by itself, i.e. to
integrate over the transverse degrees of freedom in order to identify
the stress-energy tensor eigenvalues, namely the string tension and
its energy per unit length. Let us first recall how this is done for
the Witten U(1) case by reproducing the argument of
Ref.~\cite{PPcomment}.

\par

In the U(1) situation, there is only one phase present, namely
$\alpha$, and its general solution is the same as in our case. In
fact, as discussed above, this solution is equivalent to saying that
in a small but finite neighborhood of any point $(z_0,t_0)$ on the
string, the phase can be approximated as a Taylor series $\alpha
\simeq \alpha_0 + k(z-z0) - \omega (t-t_0)+\cdots$, and since there is
an invariance of the theory under global transformations
$\alpha\to\alpha+\hbox{const}.$, it is always possible, at any given
point, to rescale $\alpha$ to the simplest solution $\alpha=kz-\omega
t$, i.e. to send $\alpha_0\to 0$.

\par

The stress-energy tensor, again for the U(1) case, does not
explicitly depend on the phase itself, but on its gradients
$\partial_\mu \alpha$, which, locally, can always be taken as
constants. As a result, the stress-energy tensor is a function of the
radial distance only if cylindrical symmetry is assumed, and its
conservation $\nabla_\mu T^{\mu\nu}=0$ implies, for $\nu=r$,
$$
\int r\,\dd r\, \left( T^r_r + T^\theta_\theta \right) = 0.
$$
This sum of terms is the same as $T^x_x+T^y_y$, and the symmetry
around the vortex also implies that both these two terms are the same,
as the choice of directions for the axis $x$ and $y$ is irrelevant,
nothing depending on the angle $\theta$. Therefore, the transverse
components of the stress tensor vanish. On the other hand, the $\nu=z$
and $\nu=t$ components of the conservation equation imply that the
mixed parts $T_{rz}$ and $T_{rt}$ both behave as $r^{-1}$, which is
not possible if this tensor is to be finite: one must impose
$T_{rz}=T_{rt}=0$. There remain the internal components $T_{ab}$ with
$a,b=z,t$: upon integration and diagonalization, they provide the
relevant functions of the state parameter $w=k^2-\omega^2$ known as
energy per unit length and tension.

\par

Unfortunately, the above does not generalize easily to the more
complicated nonabelian situation. Indeed, for the simplest possible
SU(2) case we have discussed until now, the general form of the
stress-energy tensor reads
\begin{equation}
T_{\mu\nu} = t_{\mu\nu}\left(r\right)
+\sigma^2\left[ s_{\mu\nu}(z,t)-\frac12 s^\alpha_{\ \alpha} g_{\mu\nu}\right],
\label{tmn}
\end{equation}
where
\begin{equation}
s_{\mu\nu}= \partial_\mu
 \alpha \partial_\nu \alpha +
\sin^2\alpha\left( \partial_\mu \beta \partial_\nu \beta + \sin^2
\beta \partial_\mu \gamma \partial_\nu \gamma\right)
\label{s}
\end{equation}
shows an explicit dependence in the wordsheet coordinates
and the first part $t_{\mu\nu}$ only depends on $r$. Let
us see how the above argument fails in this case.

\par

The conservation equation, as given above, with $\nu=r$, now
transforms into
\begin{equation}
\int r\,\dd r\, \left( T^r_r + T^\theta_\theta \right) = \int r^2\,\dd
r \, \left( \partial_t T_{tr} -\partial_z T_{zr} \right),
\label{stressgen}
\end{equation}
while the $z$ and $t$ components respectively give
$$
\left(\frac{\partial}{\partial r} +\frac{1}{r} \right)T_{rz} = \partial_t T_{tz}
- \partial_z T_{zz},
$$
and
$$
\left(\frac{\partial}{\partial r} +\frac{1}{r} \right)T_{tz} = \partial_t T_{tt}
- \partial_z T_{zt}.
$$
Assuming the separated form $T_{rz} = Z(r) \tilde{T}_{rz}$ and $T_{rt}
= T(r) \tilde{T}_{rt}$, with $\tilde{T}$ being independent of $r$, we
find, upon integration over $r$ of these two relations, that provided
the functions $Z$ and $T$ decay faster than $r^{-1}$, the
surface stress tensor 
\begin{equation}
\tilde{T}_{ab} \equiv \int r\,\dd r\dd\theta \, T_{ab}
\label{surface}
\end{equation}
is conserved, i.e. $\nabla_a \tilde{T}^{ab} =0$.

\par

The tensor (\ref{surface}) will contain all the relevant information
for the dynamics of the string worldsheet provided the r.h.s. of
Eq.~(\ref{stressgen}) vanishes, and this gives a necessary condition
for a two-dimensional worldsheet description to be valid. Given the
form (\ref{tmn}) of the stress tensor for the nonabelian case, it is
far from obvious that the two dimensional stress energy tensor is
automatically conserved. We shall see later that the condition that
Eq.~(\ref{stressgen}) vanishes provides a constraint on the second
time and space derivative of the angular functions $\alpha$, $\beta$
and $\gamma$.

\par

Let us now return to the crooked string in the ultralocal regime. The
surface stress energy tensor takes the form
\begin{equation}
\bar{T}^{a}_{\ b} = \left( \begin{array}{cc}
T^t_{\ t} & T^t_{\ z} \cr T^z_{\ t}  &T^z_{\ \ z} 
\end{array}\right) = \left( \begin{array}{cc}
-A+B & C \cr -C  &-A-B 
\end{array}\right),
\label{2DT}
\end{equation}
where
\begin{widetext}
\begin{equation}
A = 2\pi\int r\,\dd r\, \left\{ \varphi'^2 + \frac{Q'^2}{2q^2 r^2} +\frac12
\sigma'^2 + \frac{Q^2\varphi^2}{r^2} + \frac12\sigma^2 \left[\alpha'^2
 +\sin^2\alpha\left( \beta'^2 + \sin^2\beta
   \gamma'^2\right)\right]\right\},
\label{A}
\end{equation}
\end{widetext}
while
\begin{equation}
B=\sum_{i=\alpha,\beta,\gamma} \left(k_i^2+\omega_i^2\right) I^i
\label{B}
\end{equation}
and
\begin{equation}
C=2 \sum_{i=\alpha,\beta,\gamma} k_i\omega_i I^i
\label{C}
\end{equation}
are expressible in terms of the profile integrals
\begin{eqnarray}
I^\alpha&=&\pi\int\sigma^2 r\,\dd r, \\ \label{Ia}
I^\beta&=&\pi\int\sigma^2 \sin^2 \alpha r\,\dd r,\\ \label{Ib}
I^\gamma&=&\pi\int\sigma^2 \sin^2 \alpha \sin^2 \beta r\,\dd r .\label{Ig}
\end{eqnarray}

The energy per unit length $U$ and the tension $T$ are then obtained
as the respectively timelike and spacelike eigenvalues of this stress
tensor, namely
\begin{equation}
U = A+\sqrt{B^2-C^2} \ \ \ \ \hbox{and} \ \ \ \ \ T=A-\sqrt{B^2-C^2},
\label{UT}
\end{equation}
where the quantity $B^2-C^2$ can be expressed in terms of all the
possible Lorentz invariant scalars made from the phase gradients,
namely the parameter matrix
\begin{equation}
w_{ij} = k_i k_j - \omega_i \omega_j,
\label{wij}
\end{equation}
and we find
\begin{equation}
B^2-C^2 = \sum_{i,j=\alpha,\beta,\gamma} I^i I^j \left(2w_{ij} - w_i w_j\right),
\label{B2C2}
\end{equation}
which generalizes the abelian case.

\par

Eqs.~(\ref{UT}) and (\ref{B2C2}) show that the energy per unit length
and tension of the nonabelian current carrying string depend
explicitly on all the possible two-dimensional (worldsheet) Lorentz
invariant parameters that can be constructed out of the phase
gradients of the angular variables $\alpha$, $\beta$ and
$\gamma$. Although this induces a tremendous level of complexity for
the description of the dynamics of the string worldsheet itself, this
is however not the end of the story, for the field equations for the
angle profiles actually show another dependence, implicit this time:
under the assumption of ultralocality, the Euler equations for
$\alpha$, $\beta$ and $\gamma$, namely Eqs.~(\ref{alphadyn}),
(\ref{beta}) and (\ref{gamma}), contain the parameters
$\partial_{zz}\alpha^0-\partial_{tt}\alpha^0$,
$\partial_{zz}\beta^0-\partial_{tt}\beta^0$ and
$\partial_{zz}\gamma^0-\partial_{tt}\gamma^0$, i.e. again, all the
possible string Lorentz invariant second order derivatives. This makes
a difference with the abelian case for which, as we showed in
Sec.~\ref{ab}, these second derivatives do not enter, at any
level. Here, since they enter in the profiles, the energy per unit
length and tension indirectly depend on their values. Thus, going from
U$(1)$ to SU$(2)$, one increases the number of free parameters from
one to eight or nine, depending on whether one considers or not yet
another constraint, which we now discuss.

\par

In Sec.~\ref{secTmn}, we showed that the two dimensional stress energy
tensor is conserved only provided the r.h.s of Eq.~(\ref{stressgen})
vanishes. This, given the form (\ref{tmn}), can be implemented in two
ways. The first possibility is to simply assume the ultralocal
approximation in the stress energy tensor itself, which amounts to
saying that $s_{\mu\nu}$ in Eq.~(\ref{s}) does, in fact, depend on
neither $z$ nor $t$; in this case, $T_{\mu\nu}$ is merely a function
of the radial variable and the analysis of \cite{PPcomment} applies.

\par

Another way to impose the surface stress energy tensor to be conserved
is by expliciting the condition
\begin{equation}
\partial_t\int r^2\,\dd r\, s_{tr} = \partial_z \int r^2\,\dd r\, s_{zr} 
\label{condTab}
\end{equation}
using the expansion (\ref{alcroo}), and only then take the ultralocal
limit. This method gives a relationship between the second derivatives
of the angular variables and their gradients, hence reducing the
number of free parameters by one unit.

\par

Finally, one can use the stress energy tensor here derived to recover
the biabelian situation, which will allow to illustrate a difference
between many abelian and nonabelian currents. The U$(1)\times$U$(1)$
case of Sec.~\ref{biabelian} is obtained in the ultralocal limit by
writing $\alpha\to \alpha(r) + k_\alpha z -\omega_\alpha t$, $\beta
\to \frac{\pi}{2} + k_\beta z -\omega_\beta t$ and $\gamma\to k_\gamma
z -\omega_\gamma t$, and then assuming $t,z\to 0$. Then the stress
energy tensor above is unchanged, with now $I^\beta = I^\gamma = I_1=
\pi\int \sigma_1^2(r) \, r\,\dd r$ and $I^\alpha = I_1+I_2 = \pi\int
\left[\sigma_1^2(r) + \sigma_2^2(r)\right] \, r\,\dd r$, where the
fields are defined above Eq.~(\ref{twofields}). Setting
$k_2=k_\alpha\equiv\partial_z\psi_2$,
$\omega_2=\omega_\alpha\equiv-\partial_t\psi_2$, $\bm{k}_1^2 =
k_\alpha^2+k_\beta^2+k_\gamma^2\equiv
\left(\partial_z\psi_1\right)^2$, and $\bm{\omega}_1^2 =
\omega_\alpha^2+\omega_\beta^2+\omega_\gamma^2\equiv
\left(\partial_t\psi_1\right)^2$, we diagonalize the stress energy
tensor as above [Eq.~(\ref{B2C2})] to get
\begin{eqnarray}
B^2-C^2=&&I^2_2 \left( k_2^2-\omega_2^2\right)^2 + I_1^2 \left(
 \bm{k}_1-\bm{\omega}_1\right)^2
\left(\bm{k}_1+\bm{\omega}_1\right)^2 \cr
&&+ I_1 I_2 \left[ \left(k_2-\omega_2\right)^2
 \left(\bm{k}_1+\bm{\omega}_1\right)^2 \right.\cr
&&\hskip1cm\left.+ \left(k_2+\omega_2\right)^2
 \left(\bm{k}_1-\bm{\omega}_1\right)^2\right],
\end{eqnarray}
which is of the form of Eq.~(48) of Ref.~\cite{LPXcoupled} only
provided the vectors $\bm{k}_1$ and $\bm{\omega}_1$ are colinear,
i.e. $\bm{k}_1 = k_1 \bm{u}$ and $\bm{\omega}_1 = \omega_1 \bm{u}$,
with $\bm{u}^2 = 1$. In this case, we recover indeed
\begin{equation}
B^2-C^2 = w_1 I_1^2 + w_2 I^2_2 + 2 x I_1 I_2,
\end{equation}
where $w_i=k_i^2-\omega_i^2$ and $x=k_1 k_2 - \omega_1 \omega_2$ is
the cross product. This particular choice is that which lowers the
number of arbitrary parameters to only three, as demanded by the two
abelian current case.

\par

The biabelian current case, as discussed above, has a microscopic
structure (the field profiles) that depends solely on the squared
phase gradients $w_1=\left(\partial \psi_1\right)^2$ and
$w_2=\left(\partial \psi_2\right)^2$, even though the energy per unit
length and tension also depend on the cross product $x=\partial
\psi_1\cdot\partial \psi_1$. By contrast, the nonabelian
current-carrying case involves in a non trivial way not only the
gradients $\left(\partial \alpha\right)^2$, $\left(\partial
  \beta\right)^2$ and $\left(\partial \gamma\right)^2$, but also all
the possible combinations of cross products, namely $\partial
\alpha\cdot\partial\beta$, $\partial \alpha\cdot\partial\gamma$ and
$\partial \beta\cdot\partial\gamma$; this is clear from the dynamical
equations (\ref{alphadyn}), (\ref{beta}) and (\ref{gamma}) defining
the profiles of these angles, again provided one takes the ultralocal
limit after deriving these equations.

\section{Conclusion}

Cosmic string are an almost generic prediction of most high energy
theories, and they can have many observational cosmological
consequences. They can also be current-carrying, and this property
changes their dynamics drastically, as it has been argued that a
network of current-carrying cosmic string could overproduce
equilibrium loop configurations which, if stable, would overclose the
Universe; such strings are clearly ruled out. The last case that was
not yet studied is that for which the current carrier transforms
according to some representation of a nonabelian group, and this is
what has been presented above, in the particular (simplest) example of
(global) SU$(2)$. By means of such a toy model, we have been able to
derive the microscopic structure of a nonabelian current-carrying
string, and exhibit the characteristic features of its stress energy
tensor, out of which one obtains, through integration over the
transverse degrees of freedom, the energy per unit length and
tension. In principle, these quantities allow for a complete
calculation of the dynamics of the strings, hence of the motion of a
network.

\par

We have found many differences between the abelian and the nonabelian
situations. Where the abelian case involves a single state parameter,
the simplest nonabelian model here developed contains far more
parameters, namely at least 8. Besides, when the abelian current case,
even with more than one current, involves only the phase gradients of
the fields, the nonabelian case at hand exhibits implicit dependencies
in the second derivatives with respect to the worldsheet coordinates
of these phases. Those phases also acquire a profile, i.e. they must
vary between the string core and the exterior: in accordance with the
general Carter argument \cite{BCIII,BCIV}, the path followed by the
phases on the SU$(2)$ 3-sphere could not be smoothly projected onto
the worldsheet itself, the later being flat while the former being
intrinsically curved. Finally, whereas in the many current case the
eigenvalues of the stress energy tensor depend only explicitly on the
cross gradients, the microscopic structure - the profiles - depending
only on the squared gradients, in the nonabelian case the profiles,
and hence the energy per unit length and tension, depend on all the
possible two dimensional Lorentz invariants that can be built out of
the phase derivatives up to the second order.

\par

If cosmic strings were ever formed, it is quite likely that they would
be current-carrying, and in this category, since the well-tested
standard electroweak theory already contains a broken SU$(2)$ with a
Higgs field doublet as in our case \cite{lowmass}, the model we
developed here may be relevant, depending on the values of the unknown
coupling parameters. At the cosmological level, abelian current
carrying strings do intercommute in much the same way as non
conducting ones \cite{MatznerLaguna}. This is made possible because
the currents in both pieces of the colliding strings can merely add up
at the junction, being confined in the worldsheet through a linear
interaction. In the nonabelian case, it is likely that the essentially
nonlinear interaction terms would forbid such a simple readjustment of
the phases: it is to be expected that the intercommutation probability
is much lower than for ordinary strings. This, as is well known from
the superstring case \cite{Polchinski05}, can imply fundamentally
different cosmological consequences. Another reason why one would
expect intercommutation to be far less effective in the nonabelian
current-carrying case is also related to extra dimensions: in the
simplest Kaluza-Klein framework with a circular fifth dimension, the
extra angular variable plays the role of the current-carrier phase and
the equation of state can be calculated to be of the self-dual fixed
trace kind \cite{models} by projecting in the 4 dimensional base space
\cite{Carter90}; it can be conjectured that introducing many extra
dimension with a complicated structure can lead to currents sharing
many of the properties of the nonabelian ones discussed here.  The
intercommutation of nonabelian current-carrying cosmic string is
therefore an important open problem that deserves further
investigation.

\acknowledgments We thank B.~Carter for enlightening discussions.

\bibliography{references}  

\end{document}